# Lithographically Defined Zerogap Strain Sensors

*Mahsa Haddadi Moghaddam, Zhihao Wang, Daryll J.C Dalayoan, Daehwan Park, Hwanhee Kim, Sunghoon Im, Kyungbin Ji, Daeshik Kang, Bamadev Das\*, Dai Sik Kim\**

(A. B) Mahsa Haddadi Moghaddam, Zhihao Wang, Daryll J.C Dalayoan, Daehwan Park, Hwanhee Kim, Bamadev Das, E-mail: bamadevdas0@gmail.com

(A. B. C. D) Dai Sik Kim, E-mail: daisikkim@unist.ac.kr

(E) Daeshik Kang, Sunghoon Im, Kyungbin Ji

A- Department of Physics, Ulsan National Institute of Science and Technology, Ulsan 44919, Republic of Korea

B- Quantum Photonics Institute, Ulsan National University of Science and Technology, Ulsan 44919, Republic of Korea

C- Center for Angstrom Scale Electromagnetism, Ulsan National University of Science and Technology, Ulsan 44919, Republic of Korea

D- Department of Physics and Astronomy, Seoul National University, Seoul 08826, Republic of Korea

E- Department of Mechanical Engineering, Ajou University, Suwon, Republic of Korea

ABSTRACT: Metal thin films on soft polymers provide a unique opportunity for resistance-based strain sensors. A mechanical mismatch between the conductive film and the flexible substrate causes cracks to open and close, changing the electrical resistance as a function of strain. However, the very randomness of the formation, shape, length, orientation, and distance between adjacent cracks limits the sensing range as well as repeatability. Herein, we present a breakthrough: the Zerogap Strain Sensor, whereby lithography eliminates randomness and violent tearing process inherent in conventional crack sensors and allows for short periodicity between gaps with gentle sidewall contacts, critical in high strain sensing enabling operation over an unprecedently wide range. Our sensor achieves a gauge factor of over 15,000 at $\varepsilon_{ext}$=18%, the highest known value. With the uniform gaps of three-to-ten thousand nanometer widths characterized by periodicity and strain, this approach has far reaching implications for future strain sensors whose range is limited only by that of the flexible substrate, with non-violent operations that always remain below the tensile limit of the metal.

INTRODUCTION

Stretchable strain sensors capable of converting large mechanical deformation into electrical signals are promising candidates for soft and flexible devices, such as skin-mounted electronics[1-6], wearable health devices[7-11], and soft robotics[10, 11]. Most stretchable strain sensors use advanced composite materials, such as carbon nanotubes[12-15], graphene and its derivatives[16-20], metal nanowires and nanoparticles[21-23], and metal thin films.[24-30] Especially for metal thin films type sensors, recent studies reveal that the formation of spontaneous cracks in metal thin films on soft polymers, triggered by mechanical mismatch with the underlying substrates, enables high-performance crack-based strain sensors.[26-30] To alleviate the inherent randomness of the crack process that limits the sensing performance, efforts to partially orient the cracks perpendicular to the strain direction resulted in higher gauge factors but still the sensing range remains below 5%.[28-30] This is mainly due to the average distance between the cracks being in the 100 μm range limited by the material properties, which makes the gap opening too fast as a function of the strain, reaching the infinite resistance prematurely. What we need is a lithographically defined, controllable nanometer-to-micrometer gap opening without having gone-through the tearing process inherent in any cracking[31-34], while maintaining non-infinite resistance at high strain with reasonable gauge factors with desired scalability and consistency of the final samples.

In this paper, we present an advanced stretchable crack-based sensor without the conventional cracking process, a zerogap[35, 36] strain sensor (ZSS) where 3~5 nm-wide gaps in electrical contact are formed along the lithographically arranged periodic lines, which then widen with increasing strain resulting in increased electrical resistance. Our ZSS offers remarkable advances over previous methods, removing extreme forces involved in the tearing process, enabling any short period required for the high strain sensing, limited only by the lithography machine and not by the material characteristics. This is possible by transferring zerogap samples grown on rigid substrate onto PDMS. Employing this technology, we studied the sensitivity of ZSS sensors by changing periodicities from 5 μm to 500 μm. The characterization of the samples indicates a linear correlation between the periodicity and the gap widening under external stress. Thanks to this property, our sensor with a periodic gap of 5 μm (5ZSS) can operate over a wide strain range and exhibits an extremely high gauge factor of 15,000 at a strain of $13\% < \varepsilon_{ext} < 18\%$, which is due to the presence of regularly arranged long and dense parallel gaps. This wide strain coverage and high range of resistance gauge factors is unprecedented for metal-based crack sensor, where the application is limited to either low[28-30] or high[37, 38] strain range. In contrast, our ZSS has proven remarkably effective over a diverse span of applications such as pressure sensors, biomechanical dynamics, face recognition and human body movement. In addition, the proposed strain sensor shows a dynamic response of 10 Hz or more, faster than modulation reported before. [28-30]

RESULTS AND DISCUSSION

**Structure Design and Characterization. Figure 1**a illustrates the detailed fabrication scheme as follows: The first sacrificial layer (here vanadium 120 nm) and a bare Au film with a thickness of 50 nm are evaporated onto a rigid substrate (Si). Microscale arrays (the second sacrificial layer) are patterned on the Au layer using standard photolithography as a mask for the ion milling process. After etching the exposed part of the Au film, a second Au layer with the same thickness is deposited on the entire structure. To ensure good bonding, a 3 nm thick adhesive layer of titanium (Ti) was deposited immediately before both the 1$^{st}$ and 2$^{nd}$ Au layers were deposited. Chemical etching of the photoresist creates periodic arrays with lateral zerogap between the two Au layers, which are gently connected together. Polydimethylsiloxane (PDMS) is spin-coated onto the well-aligned zerogaps and then cured to form a thin and uniform elastomeric film (with a thickness of about 450 μm). In order to achieve superior performance, stretchability, and cyclic stability, (3-Mercaptopropyl) trimethoxysilane (MPTMS) is used as a precursor between Au patterns and the PDMS layer. It should be noted that the self-assembling MPTMS monolayer can enhance the adhesive interaction of PDMS with Au patterns. Finally, the zerogap-patterned Au film is separated from the rigid substrate by chemical etching of the first sacrificial layer and transferred to the elastomer substrate. **Figure 1**b shows a wafer-scale ZSS with a periodicity of 5 μm, and the inset is a typical 2 cm x 2 cm sample used for the characterization experiments. A field emission scanning electron microscopy (FE-SEM) image of the zerogap sample with a periodicity of 5 µm (5ZSS) after transfer to PDMS in unstretched position is shown in **Figure 1**c. The cross-sectional image of FE-SEM (**Figure 1**d) shows that although the two layers are perfectly connected in terms of both electrical and optical functions, they are in fact separated by 4 nm on average. The diagram in **Figure 1**e illustrates that the resistors are connected in series between parallel slits. An external strain perpendicular to the parallel slits which is defined as $\varepsilon_{ext}=(L-L_0)/L_0\times100(\%)$ ($L$ and $L_0$ are the stretched and relaxed lengths of the sample, respectively) caused the separation between two adjacent gold bars, which led to an expansion of the gaps from the nanometer to the micrometer scale and caused a change in resistance. For 5ZSS, the gap opening is visually confirmed by optical transmission images (**Figure 2**a, top) under an optical microscope (**Figure 2**b) and FE-SEM images (**Figure 2**a, bottom) using an in situ tensile stage (**Figure 2**c) on three different $\varepsilon_{ext}$, namely relaxed (0%), medium (10%) and fully stretched (17%) (see further details of the gap opening in **Figure S1** in the Supporting Information). The robust adhesion of the Au film to the PDMS substrate enables stable strain without delamination of the thin film. **Figure 2**d, which illustrates the gap opening under external strain measured by FE-SEM for different periodicities, shows perfect linearity for smaller period ($p \leq 50$ µm) according to $w=p\varepsilon_{ext}$, where $w$ is the gap width and $p$ is the periodicity. At larger $p$, the linearity with $\varepsilon_{ext}$ is still maintained. The insets provide FE-SEM images of gap opening around 13% strain for periodicities of 5 µm, 25 µm, 50 µm, and 100 µm. With identical external strain, the gap width in samples with lower density ($w$=12 µm for $p$=100 µm) is notably larger compared with samples with higher density ($w$=0.8 µm for $p$=5 µm - 5ZSS), which corresponds to a fifteenfold difference but smaller than (100 µm)/ (5 µm) =20. This deviation from the perfect linearity (red line) is more apparent in **Figure 2**e for $p$=100, 200, and 500 µm. In cases where all gaps are uniformly elongated and parallel, samples with a higher gap density require more energy to separate the gap junctions to a throughout the whole strain range). Insets show the single gap opening in different periodicities around 13% external strain (scale bars 1 μm). **e)** The rate of change of the gap under external strain for different periodicities (the solid red line shows a linear fit with a slope of 1 and both axes are in log scale). **f)** Diffraction patterns for a 50 μm periodicity sample illuminated by a helium neon laser with 633 nm wavelength from relaxed position to 20% strain.

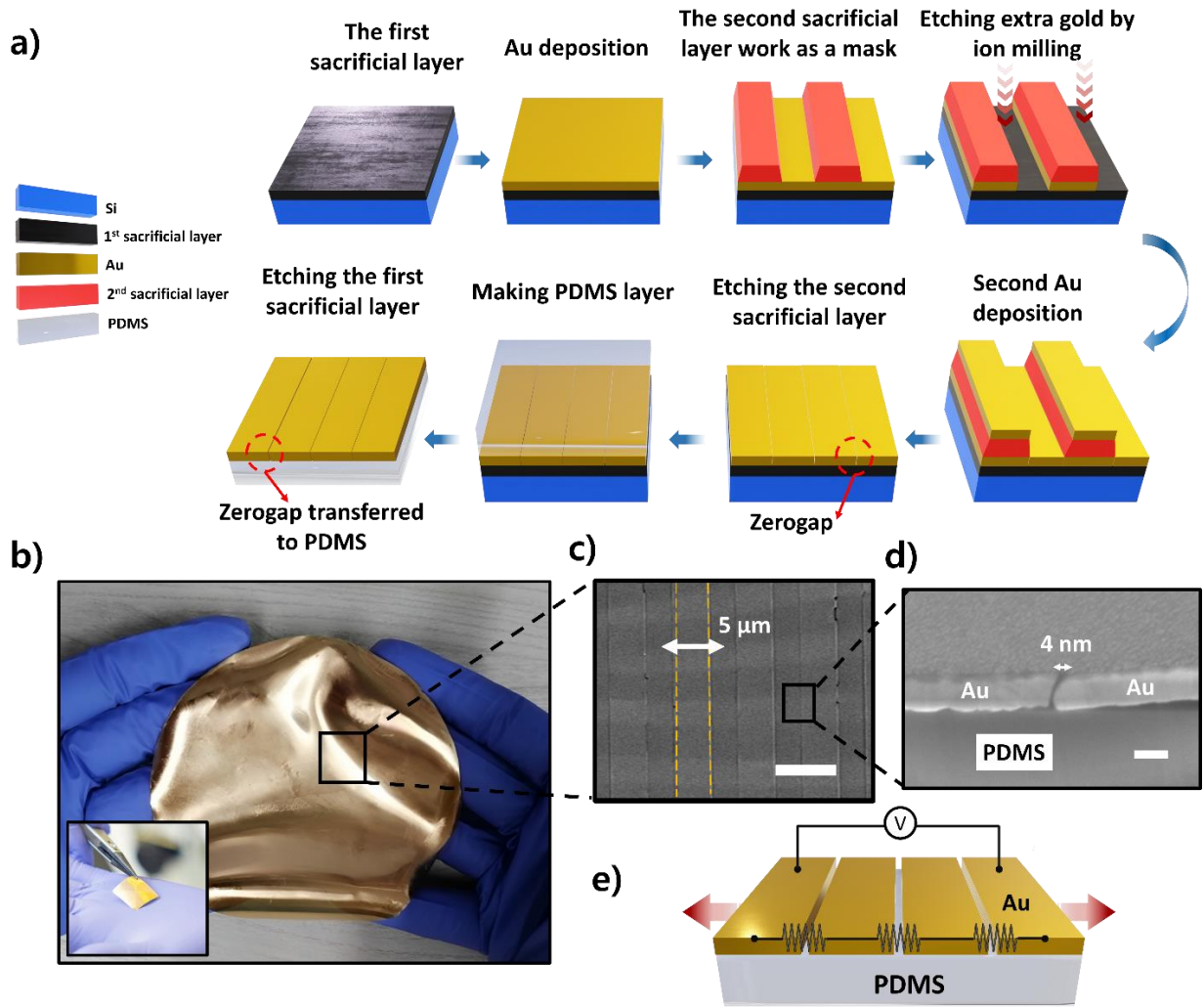

**Figure 1.** Zerogap strain sensors (ZSS) fabrication. a) the schematic diagram of the fabrication process and transferring of the zerogap embedded in micrometer periodicity to the flexible substrate. b) A photograph of the wafer scale 5ZSS (inset shows a 2 cm x 2 cm ZSS used in experiments). FE-SEM images of the c) top view and d) cross-sectional view of a fabricated ZSS sample with a periodicity of 5μm after transfer to the PDMS substrate before application of the strain (scale bars are 10 μm and 50 nm respectively). e) Schematic representation of the sensing mechanism of ZSS: changing resistance by stretching.

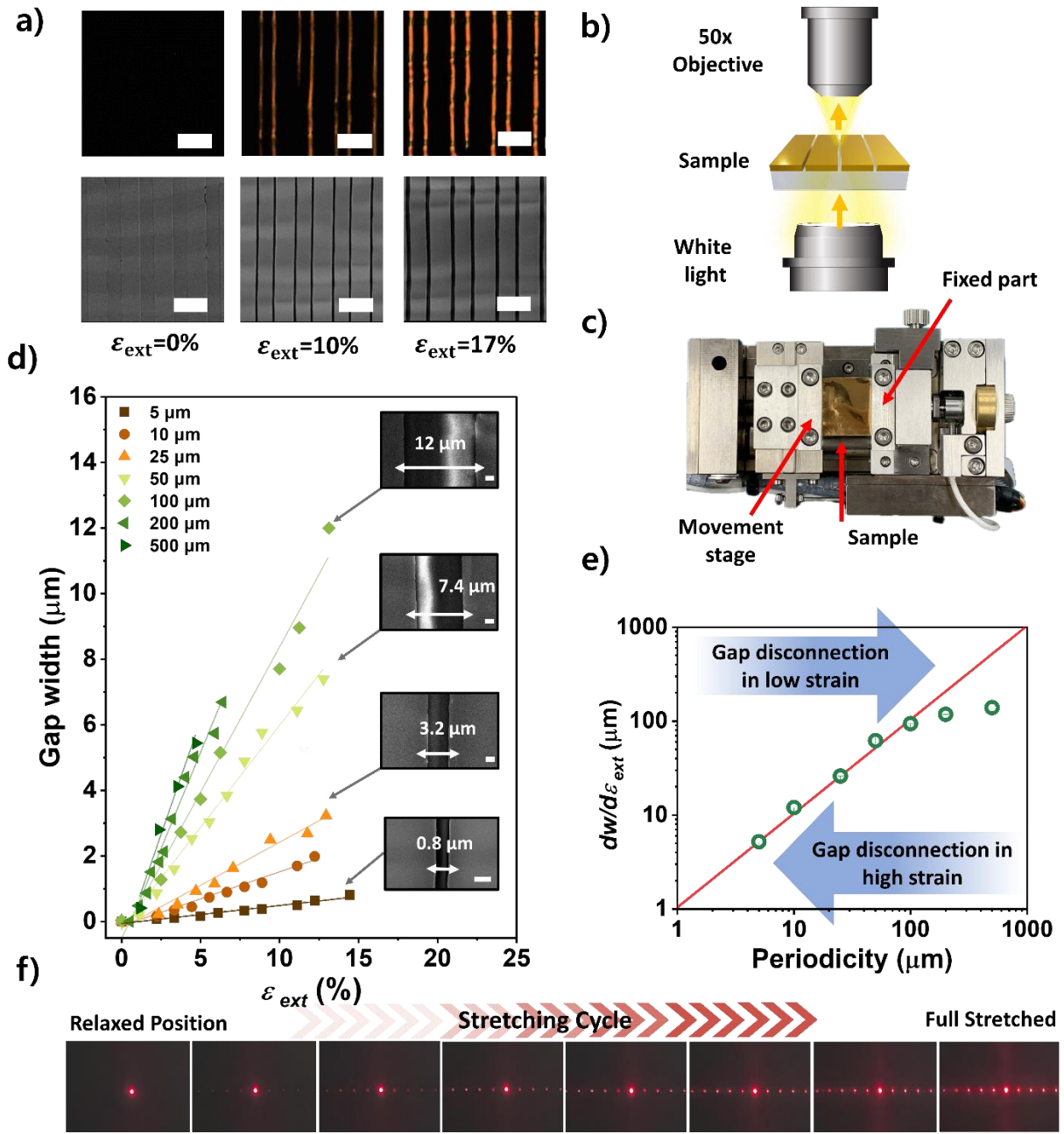

**Figure 2**. Characterization of the ZSS. **a)** Optical transmission (up) and FE-SEM images (bottom) of 5ZSS at $\varepsilon_{ext}$ =0%, 10%, and 17% (scale bars 10 μm). **b)** Schematic diagram illustrating the optical microscope setup used to analyze the transmission of the 5ZSS. **c)** A photograph of the in situ tensile stage used in FE-SEM setup. **d)** Gap width as a function of external strain for different periodicities measured by FE-SEM images. (Solid lines are linear fitting for each periodicity. The gap width exhibits a linear response

set separation, say, 1000 nm, where the resistive contact is completely lost. Most of the stress is concentrated at the periodic junctions as long as the gold bars remain crack-free (see **Figure S2**a-c in the Supporting Information) which applies to the small $p$ regime. However, for larger $p>100$ μm, inherent cracks emerge in the Au film, which effectively makes the p smaller leading to the deviation from the perfect linear line. The formation of uncontrolled cracks in large periodicities from 200 μm to 500 μm under 13% strain and comparison with Au thin films under same conditions are shown in **Figure S2** d-f in the Supporting Information. To showcase how uniform the whole sample is, diffraction patterns of the sample with an initial 50 μm periodicity is shown in **Figure 2**f when illuminated by a helium-neon laser with a wavelength of 633 nm, from a relaxed position up to an external strain of 20%. The combination of diffraction and interference effects on the light wave passing through the periodic slits produces a diffraction spectrum that appears in a symmetrical pattern on both sides of the zero-order direct light wave. Higher order diffracted wavefronts are tilted by an angle ($\theta$) according to the $sin(\theta)=n\lambda/P$, where $\lambda$ is the wavelength of the wavefront, $P=p(1+\varepsilon_{ext})$ is the periodicity of the slits and $n$ is an integer denoting the diffraction order. To comprehensively investigate the optical and electrical properties of the whole ZSS sample, 2 cm by 2 cm, we perform microwave spectroscopy and electrical measurements under various strain conditions. Transmission of ZSS with different periodicities are measured in Ku-band (≈12–18 GHz) with a vector network analyzer (see **Figure S3**a, Supporting Information), while the mechanical strain is applied using a computer- controlled piezoelectric translation stage with a 0.1 μm step (see Figure S3b, Supporting Information). **Figure 3**a shows normalized transmission amplitude for 5ZSS under varying external strain. The maximum amplitude of over 80% for $\varepsilon_{ext}$>20% implies that most gaps are open at large strains. **Figure 3**b displays transmission spectra for the 200ZSS, showing steeper rise at small strains. The sensitivity of the transmission amplitude ($\Delta t/t_0$), where $\Delta t=t-t_0$ and $t_0$ is the transmission at $\varepsilon_{ext}$=0, is plotted in **Figure 3**c at a frequency of 15 GHz for different periodicities against the external strain, consistent with the gap width dependence shown in **Figure 2**d. The performance of the ZSS is characterized by its relative resistance $\Delta R/R_0$ where $\Delta R=R-R_0$, $R_0$ and $R$ are the resistance before and after applying strain. The ZSS exhibits a linear current-voltage characteristic when subjected to different strains, indicating its proper ohmic behavior (**Figure S4** for 5ZSS in the range of $\varepsilon_{ext}$ =0-18%, Supporting Information). The analysis of samples with different periodicities, in **Figure 3**d, indicates that the 5ZSS has a wide range of sensitivities varying from $\Delta R/R_0$= 0.016 to $\Delta R/R_0$= 800 with strain changes from 0 to 18%. The external load increases the average relative resistance until two adjacent gold bars completely lose electrical contact, defining the endpoint of the strain range.

The quantification of the sensitivity of a strain sensor is expressed by its gauge factor (*GF*) serving as a critical benchmark, particularly in the realm of high-sensitivity detection. **Figure 3**e shows the local gauge factor of the 5ZSS, defined as the slope of the relative resistance versus the applied mechanical strain. The corresponding local gauge factors for 5ZSS are 311, 2,192, and 15,097 at strain ranges of 0-5%, 6-12%, and 13-18%, respectively, demonstrating the versatility of the 5ZSS in all desired strain ranges compared to other metal-based strain sensors. The inset show FE-SEM images of the cross-section of the gaps in slightly stretched, moderately stretched and fully opened

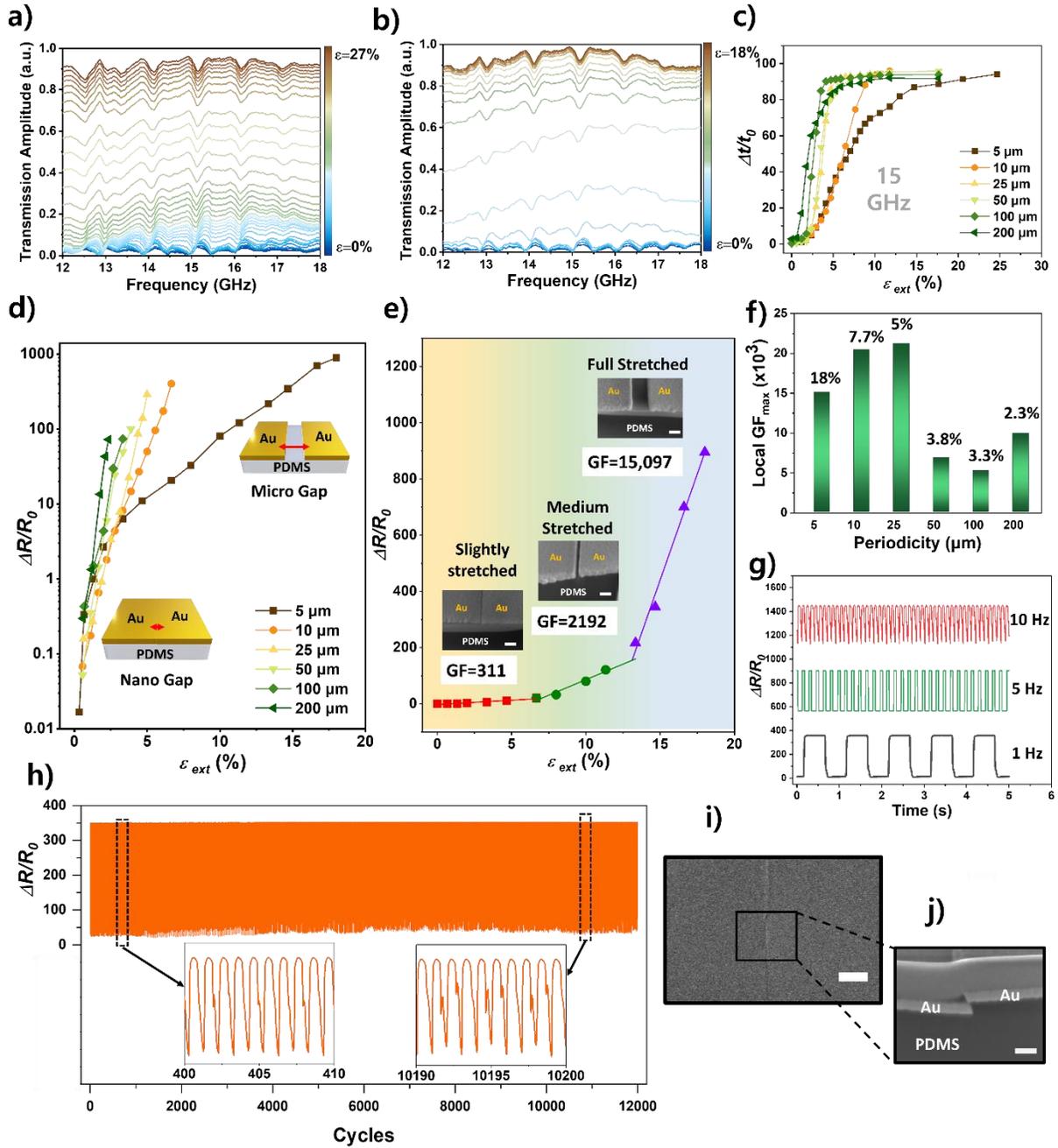

**Figure 3.** Normalized transmission spectra for **a)** 5ZSS and **b)** 200ZSS for various external strains in the microwave regime with a frequency range of 12 to 18 GHz. **c)** Amplitude transmission change ($\Delta t/t_0$) at 15 GHz for different periodicities against different external strain ($\varepsilon_{ext}$). **d)** Relative resistance-strain curve of different periodicities. (the axis of $\Delta R/R_0$ is in log scale). **e)**

Sensing curves for 5ZSS and related local gauge factor at indifferent strain range. The inset FE-SEM images show cross sections of the gap for 5ZSS in relaxed position, medium strain and full stretched (scale bars are 500 nm). **f)** The maximum gauge factor for samples with different periodicities. **g)** Cyclic responses under external strain at modulation of 1 to 10 Hz. **h)** 10 Hz durability test of 5ZSS for 12000 cycles under 15% strain. Insets represent enlarged views of 10 cycles. The FE-SEM images of **i)** top view and **j)** cross sectional view of the 5ZSS gap after fatigue test (scale bars 1 μm and 100 nm, respectively).

state. The local gauge factor for different periodicities is displayed in **Figure S5** a-e (Supporting Information) and reveals the sensitivity at indifferent strain range. The comparison of the maximum gauge factor for different periodicities at their maximum strain limit is shown in **Figure 3**f. The highest gauge factor measured is 21,194 for 25ZSS at a strain range of 5%. Compared to other recently introduced metal-film strain sensors, our nano-to-micro strain sensors show superior performance in terms of sensitivity and the gauge factor limit. We compared the sensing performance of our ZSS with other thin film strain sensors in **Table S6** in the Supporting Information. We also measured the point gauge factors related to the sensitivity of the applied strain, defined as $GF=\Delta R/\varepsilon_{ext}R_0$, as shown in **Figure S7** (Supporting Information). A gauge factor of over 6038 is achieved for 10ZSS at 6.7% strain and 4971 for 5ZSS at 18% strain before the samples completely lose the electrical connection. As a proof of the ultra-high gauge factor of ZSS, we prepared a thin gold layer with the same thickness without pattern on the PDMS substrate. We used the same method for the binding of gold and PDMS: MPTMS molecules and generated cracks by initial stretching. We compared the performance of the thin film crack sensor by performing all experiments and measurements under the same conditions as the ZSS. **Figure S8**a in the Supporting Information compares the change in resistance of the sample with the 5ZSS and the thin film. The increase in resistance observed in the thin film when subjected to an external strain is much slower and with a much lower gauge factor than in 5ZSS (**Figure S8**b, Supporting Information). This phenomenon is attributed to the limited propagation of random microcracks along the surface of the thin film, which prevents their extensive opening and subsequent loss of electrical contact. In addition to the merit of broad sensing range with high gauge factor, these devices displayed significant durability throughout stretch and release tests, without any discernible degradation in performance. We analyzed the performance of our 5ZSS sensor at a frequency of 0.1 Hz at the applied strain of 0-15% in **Figures S9**a (Supporting Information). **Figure 3**g illustrates the frequency-dependent durability test of the 5ZSS sensor over a range of ultrafast modulation: 1 Hz, 5 Hz, and 10 Hz (and with smaller modulation depth for 20 Hz in Figure S9b in the Supporting Information). Importantly, the sensors consistently demonstrate a stable change in resistance, regardless of the speed at which the external load is applied. This phenomenon is due to the strong bonding between the gold patterns and PDMS, which guarantees the capability of our zerogap strain sensor to withstand ultra-high frequency modulations, a problem often faced by other strain sensors.[26] The remarkably fast response time of 11 ms for strain and 32 ms for recovery at 10 Hz is well suited for fast and accurate sensor measurements, as demonstrated by the change in resistance in Figures S9c and S9d in the Supporting Information for high (10 Hz) and low (0.1 Hz) modulation, respectively. In addition, the robustness of the sensor was thoroughly tested by showing negligible changes in the 10 Hz duration of the relative resistance response for 12,000 cycles in the 0-15% strain (**Figure 3**h), showing identical peak patterns throughout. The FE-SEM images in top view (**Figure 3**i) and in cross section (**Figure 3**j) show that the gaps are still well connected even after modulation, except that some parts show a slight overlap.

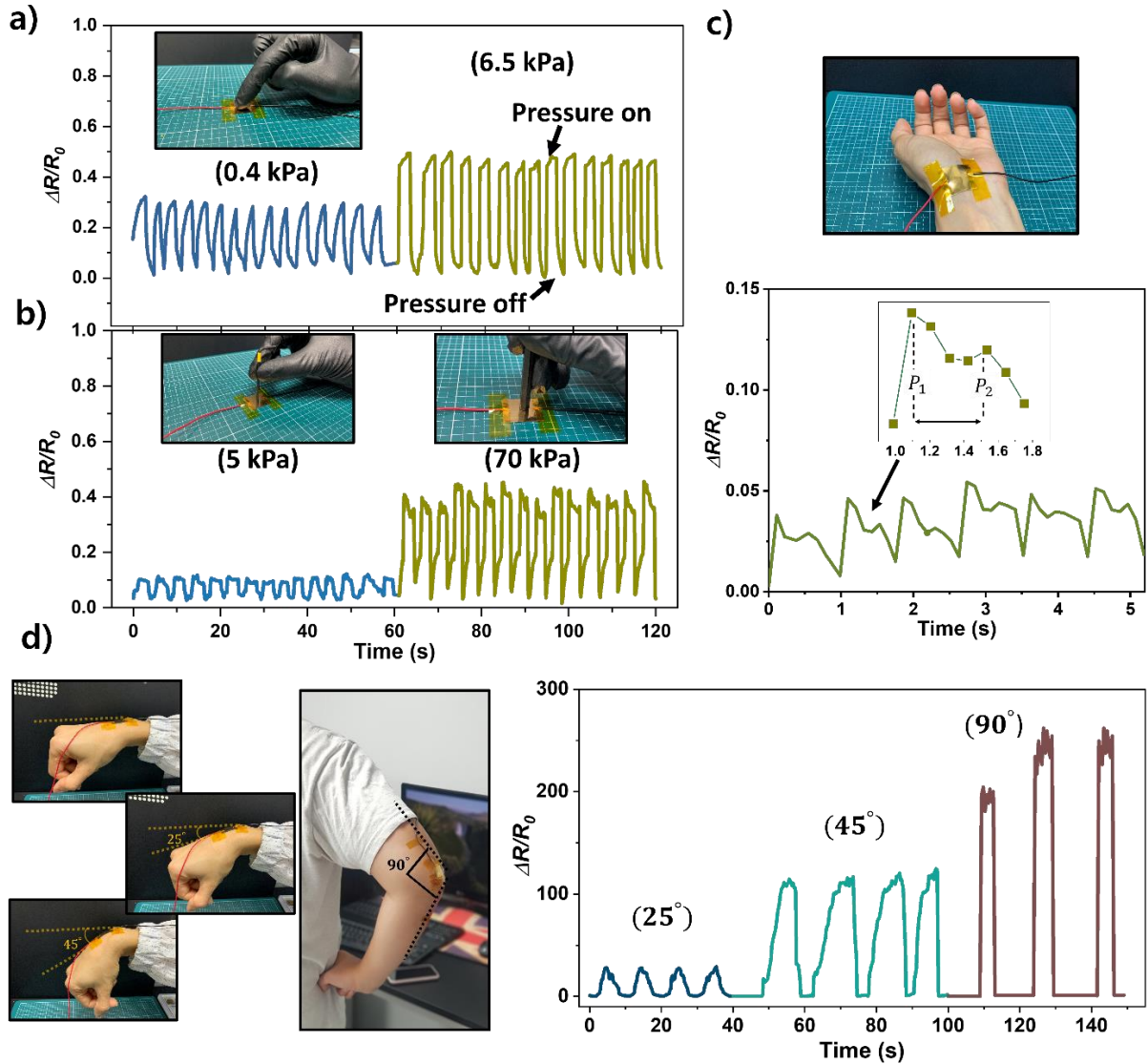

**Figure 4.** Detection of mechanical stimuli of the 5ZSS by **a)** a slight finger touch with a pressure of 0.4 kPa and a hard pressing of the finger with 6.5 kPa, **b)** pressing of hard objects with different surface areas (5 kPa and 70 kPa). **c)** Pulses of the radial artery with the sample attached to the wrist under normal conditions. The inset shows one enlarged single beat. **d)** Detection of hand movements by attaching the sensors to the wrist with different angles of θ=25˚ and θ=45˚and to the arm with θ=90˚.

**Wide strain range applications of 5ZSS.** Due to their exceptional adjustability, remarkable sensitivity, high strain capability, robust reliability, fast response times, and ease of fabrication, 5ZSS can be useful for an impressive array of applications, such as pressure sensing[39-43], human-machine interfaces and wearable technologies.[44-46] The "Experiments section" describes in detail the specific techniques for attaching the sensors to the human skin and the applied measurement methods. In the following, we focus on 5ZSS, which has the widest strain range before becoming insulating, owing to its small period. To demonstrate that 5ZSS is suited for various low and high strain applications, we present the broad applicability of our sensor by recording the pressure strain and muscle movements of the human body. We have carried out tests with our ZSS as a wide-range pressure sensor. The strain signals recorded in real time show the cycles of strain and relief caused by the external pressure being turned on or off. The sensor consistently demonstrates an elastic and stable response across different pressure levels, including a slight finger touch at 0.4 kPa, finger pressure at 6.5 kPa, narrow metal tip at 5 kPa, and thick metal tip at 70 kPa each corresponding to different contact areas (see **Figures 4**a and **4**b). As a very low strain application, we used the sensor to determine the radial artery pulse in a healthy volunteer by attaching the 5ZSS to the subject's wrist (**Figure 4**c). The blood pulse shows distinct systolic ($P_1$) and diastolic ($P_2$) peaks separated by 0.44s and an artery augmentation index $AI_r=P_2/P_1=0.71$ within a healthy range for a 40-year-old.[47] The same 5ZSS can be employed to detect significant deformations during complicated human movements, such as muscle contractions demonstrating the capability for large strain detection. The 5ZSS was attached to the wrist and reliably detected muscle movements that occur during flexion. The sensor was subjected to tensile forces resulting in changes in resistance

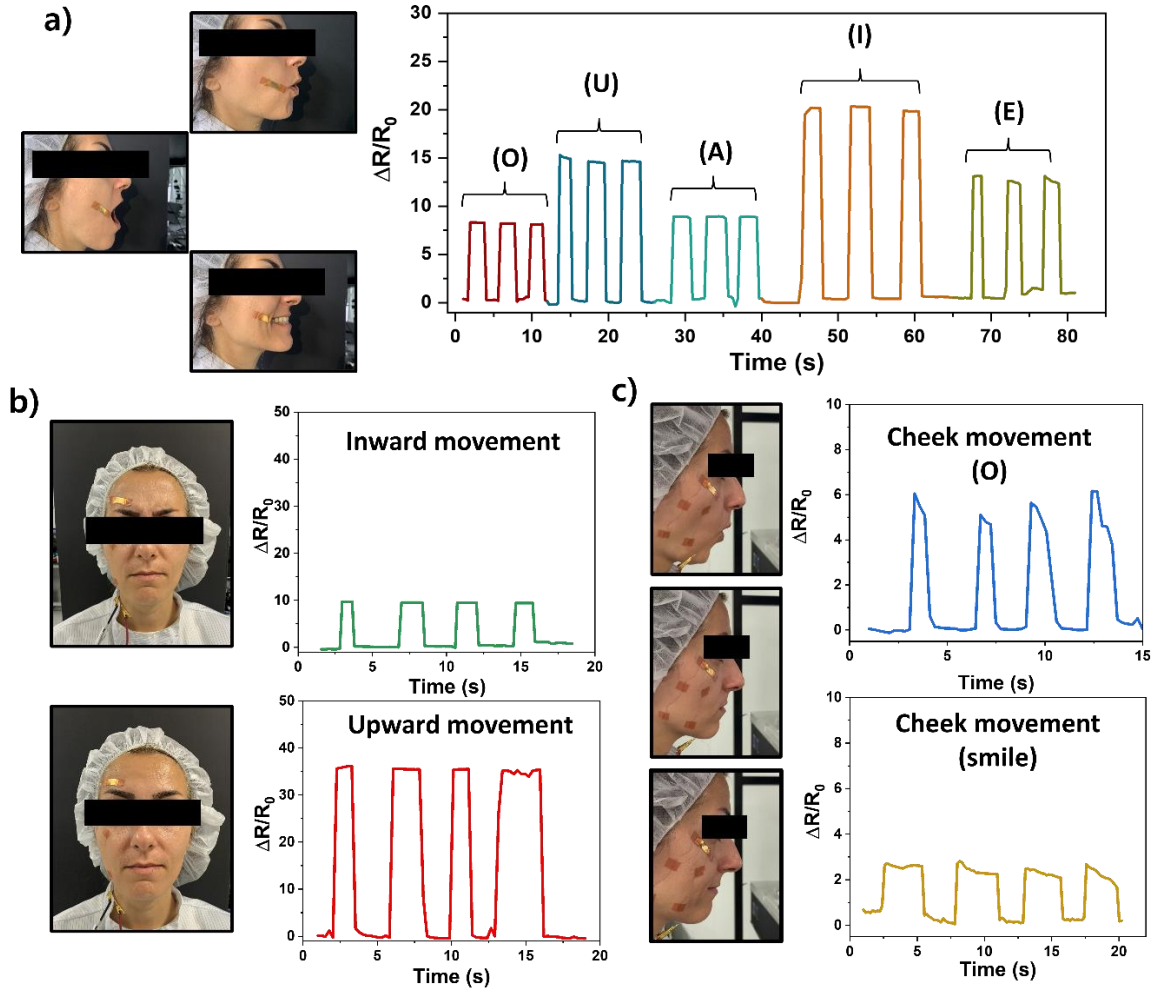

**Figure 5.** a) Images of the 5ZSS sensor attached to the corner of mouth and extracted strain signal during the vocalization of "O", "U", "A", "I" and "E". The response pattern of the 5ZSS attached to b) the upper edge of the eyebrows to recognize the inward and upward movement and c) the cheeks for the expression of "wonder" and "smile".

while the wrist was moved at different angles namely $\theta=25°$ ($\Delta R/R_0=30$) and $\theta=45°$ ($\Delta R/R_0=140$) (**Figure 4**d). It was also attached to the arm to detect $\theta=90°$ movements, resulting in a deformation of $\Delta R/R0=250$, demonstrating its ability to detect significant strains. The achieved sensitivity, spanning a remarkable range of $\Delta R/R0$ from 0.05 to 250, surpasses the detection capabilities of traditional metal film strain sensors.[26, 28, 29, 31]

Another application that requires high sensitivity for 5-10% of strain range is the recording of signals from facial muscle movements. Earlier studies identified vowels that are essential units in speech, and emotions such as smiling or anger, both through facial skin movements.[48-50] Our 5ZSS was applied to three areas of the face prone to experiencing most strain, to measure vowel speech and facial expressions. The signals from the mouth were recorded in real time during the vocalization of "O", "U", "A", "I" and "E". **Figure 5**a shows the photograph of the subject's face with sensors attached to the corner of the mouth and the stress signals during the utterance of the five different vowels, repeated three times for each vowel. The sensors were also attached to the upper edge of the eyebrows (see **Figure 5**b) to detect inward and upward movement of the eyebrow for "anger" and "wonder", respectively, and also to the cheeks (see **Figure 5**c) **for** different expressions such as "wonder" and "smile". The sensitivity of 5ZSS allows distinguishing all the five vowels characteristic as well as characteristic facial expressions ranging $2<\Delta R/R_0<40$, much better than conventional face recognition research limited a much smaller range, $\Delta R/R_0<10$.[25, 45, 50] Our strategy enables high performance designs and shows significant potential for practical applications in implantable soft electrodes, electrophysiological signals on the body surface and smart sensors, paving the way for the advancement of future robotics.

CONCLUSIONS

We lithographically defined periodic zerogaps on PDMS to be widened with strain Unlike random cracks in conventional metal thin films, our zerogap strain sensor achieves substantial gap enlargement from nanometer to micrometer scale with uniformity and repeatability. At the smallest periodicity of 5 μm, we observed a large gauge factor at the highest strain of 18%, demonstrating that we can cover a wider range of strain by simply reducing the periodicity. Our proposed strain sensor shows an impressive dynamic response in both speed and robustness in detecting fast modulations. This breakthrough in performance underlines the immense potential of our strain sensor for applications that require wide range stretchability.

MATERIALS AND METHODS

**Fabrication of ZSS**: As shown schematically in Figure 1a, 120 nm vanadium as the first sacrificial layer, 3 nm titanium as the adhesion layer and 50 nm gold are deposited on a 500 um thick silicone substrate using an e-beam evaporation system (KVE-E2000, Korea Vacuum Tech). To create a pattern on the Au layer, the process starts with coating AZ5214E photoresist with a spin coater at 4000 rpm for 60 seconds. Subsequently, the sample covered with a photomask with a 1:1 line/space stripe pattern (with different periodicity) is exposed to UV light at a wavelength of 365 nm in a mask aligner (MDA-400S, MIDAS). The desired photoresist pattern on the Au film is obtained after immersing the sample in AZ 300 MIF developer for 50 seconds. The underlying unprotected Au layers with the photoresist pattern are sequentially milled (KVET-IM4000, Korea Vacuum Tech) for 44 seconds using an argon ion beam with an incidence angle of 0°. After milling, a 3 nm thick titanium adhesion layer is applied, followed by a 50 nm thick gold secondary layer. The final step is to immerse the sample in NMP solvent at 90°C for 3 hours to remove the photoresist layer, resulting in the zerogap structure. To generate MPTMS layers in liquid form, 100 μl MPTMS was mixed with 100 ml ethanol to achieve desired concentrations. Subsequently, the Si substrates with zerogap pattern were immersed in a Petri dish containing 10 ml of the MPTMS solution for 30 minutes and then dried with N2. The MPTMS treated sample was coated with the mixture of 1:10 PDMS base and curing agent at 500 rpm for 60 s and cured in an oven at 80° for 1 h. To achieve a desired thickness of 450 μm, the PDMS coating process is repeated twice. Finally, by immersing the sample in a chromium etchant, the vanadium etched away, resulting in the detachment of the zerogap structure from the silicon substrate, allowing it to transfer onto the PDMS.

**FE-SEM measurement:** The FE-SEM system (JSM-F100, Jeol), which is equipped with an in situ tensile stage, is used for direct observation of gap modulation under different strain conditions. The dual-beam FIB system (Helios Nano Lab 450) is used for the cross sectional images of the zerogap.

**Microwave measurements:** In the Ku-band frequency range (12-18 GHz), the Agilent Technologies E5063A ENA Series Network Analyzer was used, employing a vertical waveguide setup (see Figure S3a in the Supporting Information) with the sample in the center. Microwave measurements were performed with a pair of open rectangular waveguides (62EWGN) connected to the network analyzer. The specific aperture size for the Ku-band was 15.80 mm x 7.90 mm to support the TE10 mode in this frequency range.

**Preparation of the ZSS for different application:** A wafer-scale ZSS with a periodicity of 5 μm was fabricated on PDMS and cut with a paper cutter into different sizes such as (2 cm × 2 cm) and (1 cm × 2 cm) for different applications and all were subjected to a tenfold pre-stretching process before use.

**Electrical connection:** The copper wires were attached to the sensor and connected to the source meter, (Keithley 2450). The voltage 1 V was used as the input signal and the measured current as the output signal. The strain signals were analyzed in the form of resistance during stretching and releasing are studied in a real-time measurement in time length (number of time steps).

*Pressure sensing*: The 2 cm × 2 cm size 5ZSS sensors were attached with Captone adhesive tapes to the flat surface of the weighting machine with an accuracy of 0.01 g. We have measured the contact area of the pressure for the objects (here different sizes of screwdrivers as metallic objects and finger) and calculated Pressure=(Force)/ (Contact Area). The initial resistance of the sensor is approximately $R_0$=4 Ω and slightly varied for different samples. The maximum resistance experienced during the 70 kPa was 6.5 Ω, $\Delta R/R0$~ 0.5, corresponding to the strain of 1%. The total free sensor length L (the length not limited by the contact tapes) was 18 mm, so the total maximum strain extension was $L_0 = L\varepsilon_{ext}/100 = 18 \times 1/100 = 0.18$ mm.

**The radial artery pulse:** To cover a larger area of the wrist to detect the pulse, the 2 cm × 2 cm 5ZSS sensor was attached to the skin of the wrist with adhesive strips. The initial resistance of the pulse was 23.4 Ω, and the maximum resistances for peaks *P1* and *P2* were 24.75 Ω and 24.36 Ω respectively.

**Hand and arm motion with different angles:** The 2 cm × 2 cm 5ZSS sensors were attached to the wrist and arm to record hand movements at different angles. For a movement of 25°, the initial resistance was $R_0$=12 Ω and reached the maximum resistance *R*=346 Ω, $\Delta R/R0$~30, corresponding to a stretch of 8% and a maximum total stretch of $L_0$=1.44 mm. A movement of 45° with $R_0$=20 Ω reaches the maximum resistance *R*=2831 Ω, $\Delta R/R0$~140, ε=11% and a maximum strain of $L_0$=2 mm. Finally, the movement of 90° with $R_0$=16 Ω reaches the maximum resistance *R*=4010 Ω, $\Delta R/R0$~250, ε=15% and the maximum strain of $L_0$=2.7 mm.

**Face gesture and vowels identification:** The 1 cm × 2 cm 5ZSS were placed on the different parts of the face, especially on the areas that are subject to particularly high stress. In order to recognize the different pronunciation of vowels, the sensor was placed near the mouth and detect "O" with $\Delta R/R0$=(1532 Ω -166 Ω)/(166 Ω) =8.2, "U" with $\Delta R/R0$=(782 Ω -50 Ω)/(115 Ω)=14.6, "A" with $\Delta R/R0$=(1653 Ω -166 Ω)/(166 Ω)=8.9, "I" with $\Delta R/R0$=(2346 Ω -110 Ω)/(110 Ω)=20.3, and "E" with $\Delta R/R0$=(1837 Ω -130 Ω)/(130 Ω)=13.1.

The sensor is placed on top of the eyebrow and detect "anger" and "wonder" expression by a resistance change in the range of $\Delta R/R0$= (1209 Ω -115 Ω)/(115 Ω) =9.5 and $\Delta R/R0$= (7300 Ω -200 Ω)/(200 Ω)=35.5, respectively. For cheek movements, the sensor detect the expression "wonder" and "smile" with Δ*R/R0*=(500 Ω -78 Ω)/(78 Ω)=5.4 and Δ*R/R0*=(160 Ω -46 Ω)/(46 Ω)=2.47, respectively.

**Informed consent statement**

The described tests on humans do not require institutional review board (IRB) approval because these experiments do not affect the body or physiology of living humans, and all participants are authors of this paper who participated after giving informed consent.

ASSOCIATED CONTENT

**Supporting Information**. The Supporting Information is available free of charge.

FE-SEM images and optical transmission images of the whole stretching process, FE-SEM images of different periodicities and the thin film, schematic diagram of the microwave transmission setup, current−voltage curves of 5ZSS, resistance sensing curves for different periodicities, comparison of the sensing performance of ZSS with other thin film strain sensors, point gauge factor for different periodicities, the comparison of sensitivity of ZSS and thin film, gap modulation of 5ZSS.

AUTHOR INFORMATION

**Corresponding Authors**

* Dai Sik Kim, E-mail: daisikkim@unist.ac.kr

https://orcid.org/0000-0001-8269-1340

- Department of Physics, Ulsan National Institute of Science and Technology, Ulsan 44919, Republic of Korea
- Quantum Photonics Institute, Ulsan National University of Science and Technology, Ulsan 44919, Republic of Korea

- Center for Angstrom Scale Electromagnetism, Ulsan National University of Science and Technology, Ulsan 44919, Republic of Korea
- Department of Physics and Astronomy, Seoul National University, Seoul 08826, Republic of Korea

* Bamadev Das, E-mail: bamadevdas0@gmail.com

- Department of Physics, Ulsan National Institute of Science and Technology, Ulsan 44919, Republic of Korea
- Quantum Photonics Institute, Ulsan National University of Science and Technology, Ulsan 44919, Republic of Korea

AUTHORS

Mahsa Haddadi Moghaddam, Zhihao Wang, Daryll J.C Dalayoan, Daehwan Park, Hwanhee Kim,

- Department of Physics, Ulsan National Institute of Science and Technology, Ulsan 44919, Republic of Korea
- Quantum Photonics Institute, Ulsan National University of Science and Technology, Ulsan 44919, Republic of Korea

Daeshik Kang, Sunghoon Im, Kyungbin Ji

- E- Department of Mechanical Engineering, Ajou University, Suwon, Republic of Korea

**Author Contributions**

Mahsa Haddadi Moghaddam. fabricated the samples, designed, and performed the experiments, the data analysis and wrote the manuscript; Bamadev Das. designed the experiments; Zhihao Wang, Daryll Dalayoan, Daehwan Park, Hwanhee Kim, Sunghoon Mi, and Kyungbin Ji. helped the measurements; Daeshik Kang, reviewed and edited the manuscript; Dai Sik Kim. conceived the idea, led the work, reviewed, edited, and supervised this work.

Notes

The authors declare no competing financial interest.

ACKNOWLEDGMENT

This work was supported by the National Research Foundation of Korea (NRF) grant funded by the Korean government (MSIT: NRF-2015R1A3A2031768), the National R&D Program through the National Research Foundation of Korea (NRF) funded by Ministry of Science and ICT (2022M3H4A1A04096465), the Basic Science Research Program through the National Research Foundation of Korea (NRF) funded by the Ministry of Education (NRF-2022R1I1A1A01073838), and the MSIT (Ministry of Science and ICT), Korea, under the ITRC (Information Technology Research Center) support program (IITP-2023-RS-2023-00259676) supervised by the IITP (Institute for Information & Communications Technology Planning & Evaluation).